\documentclass[apj]{emulateapj}
\usepackage{epsfig}
\usepackage{float}
\slugcomment{as of \today}
\shorttitle{Multiple Population Dynamics}
\shorttitle{Richer et al.}

\begin{document}

\title{ A Dynamical Signature of Multiple Stellar Populations in 47 Tucanae}
\author{Harvey~B.~Richer\altaffilmark{1},
  Jeremy~Heyl\altaffilmark{1}, Jay~Anderson\altaffilmark{2}, Jason~S.~ Kalirai\altaffilmark{2}, Michael~M.~ Shara\altaffilmark{3}, Aaron~Dotter\altaffilmark{4}, Gregory~G.~Fahlman\altaffilmark{5}, R.~Michael~Rich\altaffilmark{6}} 

\altaffiltext{1}{Department of Physics \& Astronomy, University of
  British Columbia, Vancouver, BC, Canada V6T 1Z1;
  richer@astro.ubc.ca, heyl@phas.ubc.ca } 
\altaffiltext{2}{ Space Telescope Science Institute,Baltimore MD
  21218; jayander@stsci.edu; jkalarai@stsci.edu }
 \altaffiltext{3}{Department of Astrophysics, American Museum of Natural
  History, Central Park West and 79th Street, New York , NY 10024;
  mshara@amnh.org} 
  \altaffiltext{4}{Research School of Astronomy and Astrophysics, The Australian National University, Weston, ACT 2611, Australia; aaron.dotter@gmail.com} 
\altaffiltext{5}{National Research Council, Herzberg Institute of
  Astrophysics, Victoria, BC, Canada V9E 2E7;
  greg.fahlman@nrc-cnrc.gc.ca} 
  \altaffiltext{6}{Division of Astronomy and Astrophysics, University of
  California at Los Angeles, Los Angeles, CA, 90095;
  rmr@astro.ucla.edu}

\begin{abstract}

  Based on the width of its main sequence, and an actual observed split when viewed through particular filters, 
  it is widely accepted that 47 Tucanae contains multiple stellar populations.  In this
  contribution, we divide the main-sequence of 47 Tuc into four color
  groups, which presumably represent stars of various chemical
  compositions. The kinematic properties of each of these groups is
  explored via proper-motions, and a strong signal
  emerges of differing proper-motion anisotropies 
  with differing main-sequence color; the bluest main-sequence stars exhibit the largest
  proper-motion anisotropy which becomes undetectable for the reddest
  stars.  In addition,  the bluest stars are also the most centrally concentrated. A simiilar analysis for SMC
  stars, which are located in the background of 47 Tuc on our frames,
  yields none of the
  anisotropy exhibited by the 47 Tuc
  stars. We discuss implications of these results for possible
  formation scenarios of the various populations.

\end{abstract}


%%%%%%%%%%%%%%%%%%%%%%
%
\section{Introduction}
\label{sec:intro}
%
%%%%%%%%%%%%%%%%%%%%%%
% 
Previous to about 1980, the general paradigm for globular star
clusters was that they were simple stellar populations, that is all
stars had uniform chemical composition and were all the same
age. However, since that time, numerous spectroscopic studies have
shown that many of these clusters exhibit chemical composition
variations among their stars likely caused by H-burning via the hot
CNO cycle (see Gratton et al. 2004 for a nice review). Recent
imaging observations with the Hubble Space Telescope (HST) that produced
exquisitely precise photometry has extended this picture to many if
not all clusters (see Piotto 2009 for a recent review).  However, what
is currently lacking is detailed insight into the manner of formation
of the various stellar populations in such a cluster. Key input could
potentially come from the observation of different dynamics or spatial
distributions of the various populations. Sollima et al. (2007) did find radial
variations among blue and red main sequence (MS) stars in $\omega$ Cen, but
this observation was not coupled with any proper motion (PM) information.
  
The MS of 47 Tuc is broader than observational error
alone can explain. This was first pointed out by Anderson et
al. (2009) and examined in some detail by Milone et al.\ (2012a) who
clearly demonstrated that the CMD width could be explained as a second-generation
population, making up 70\% of the total, enriched in both He and N and 
depleted in C and O.

Vesperini et al. (2013), using N-body simulations, have explored the
behavior of first- and second-generation stars in a
multi-population cluster. In general they find that complete mixing of
the two populations does not occur until the cluster is well advanced
dynamically; that is when it has lost upwards of 70\% of its mass due
to two-body relaxation. Giersz and Heggie (2011) performed a Monte Carlo simulation of 47 Tuc
and concluded that it has likely lost only about 45\% of its initial mass. We can thus expect that 47 Tuc's various populations will not
yet be well-mixed and that it might be possible to observe either
kinematic and/or spatial differences amongst its various stellar
populations. Indeed there already exists a wealth of observations on the radial distributions of various stellar 
populations going back over 30 years (e.g. Norris and Freeman 1979, Milone 2012a) most of which suggest that
stars exhibiting evidence for CN-enhnacement are more centrally concentrated than CN-poor stars. Coupling this with
kinematic observations could well provide critical clues to
the multiple population scenario.

In \S\ 2 below we discuss the observations relevant to the
current study and follow this in \S\ 3 with an exploration of the PM
kinematics of cluster stars yielding strong evidence for
differences amongst stars of differing MS colors (chemical
compositions). \S\ 4 presents a search for radial differences and \S\
5 discusses possible formation scenarios for the various stellar
populations based on their currently observed motions and
distributions.
  
\section{The data}
\label{sec:images}

Our team was awarded 121 HST orbits in Cycle 17 to image 47 Tuc
(GO-11677).  The main science goal was to obtain photometry with the
ACS F606W and F814W filters that would go deep enough to study the
entirety of the white dwarf cooling sequence. A detailed discussion of
the observations can be found in Kalirai et al.\
(2012). We have supplemented these observations with F606W and F814W images in the archive taken between 2002 and 2012 with ACS/WFC and WFC3/UVIS.  These images were taken at a very wide variety of offsets, orientations, and exposure times.  Sources were found in each image independently and measured with a library PSF that had been constructed from other data sets (see Anderson and King 2006, AK06).  Positions were corrected for distortion using the Bellini et al. (2011) solution for WFC3/UVIS and the AK06 solution for ACS, with an adjustment for the fact that the ACS solution has changed slightly since SM4.

We collated these many starlists in a reference frame based on that described in AK06.  We identified the member stars from the CMD and used their positions in the reference frame and the individual frames to define a linear transformation from each 2048x2048-pixel amplifier into the reference frame.  For each star, this gives us a reference-frame position for each exposure, and a linear fit can be made to this time series to compute the PMs and determine their errors.  We iterated the procedure, which allowed us to do a better job of outlier rejection in the determination of the transformations.  
The result of this process is a list of average positions and PMs (and their errors) for each star in the list. 
Since the field was observed at a wide variety of pointings and orientations over $\sim10$ years, we would expect most minor issues such as errors in the distortion solution or errors in the pixel-based CTE correction to average out to zero and introduce only random errrors, which would be reflected in our empirical PM-error estimates.  We have validated this by computing local corrections for the reference-frame position of each star in each exposure based on the 25 closest stars within $\pm 0.5$ magnitude of its brightness (as described in McLaughlin et al. 2006, and Anderson and van der Marel 2010).  This strategy should remove any residual distortion or CTE errors.  We find that the results of this local test are indistinguishable from those of our more global procedure.
 
The Small Magellanic Cloud (SMC) lies in the background of 47 Tuc. Fortunately the cluster is moving
with respect to the SMC by several milliarcseconds (mas) per year, so
that over the time span of all available images the differential
motion of the two systems amounts to nearly a whole ACS pixel (50
mas) allowing us to separate out these populations via PMs.

\begin{figure}[t]
\epsfig{file=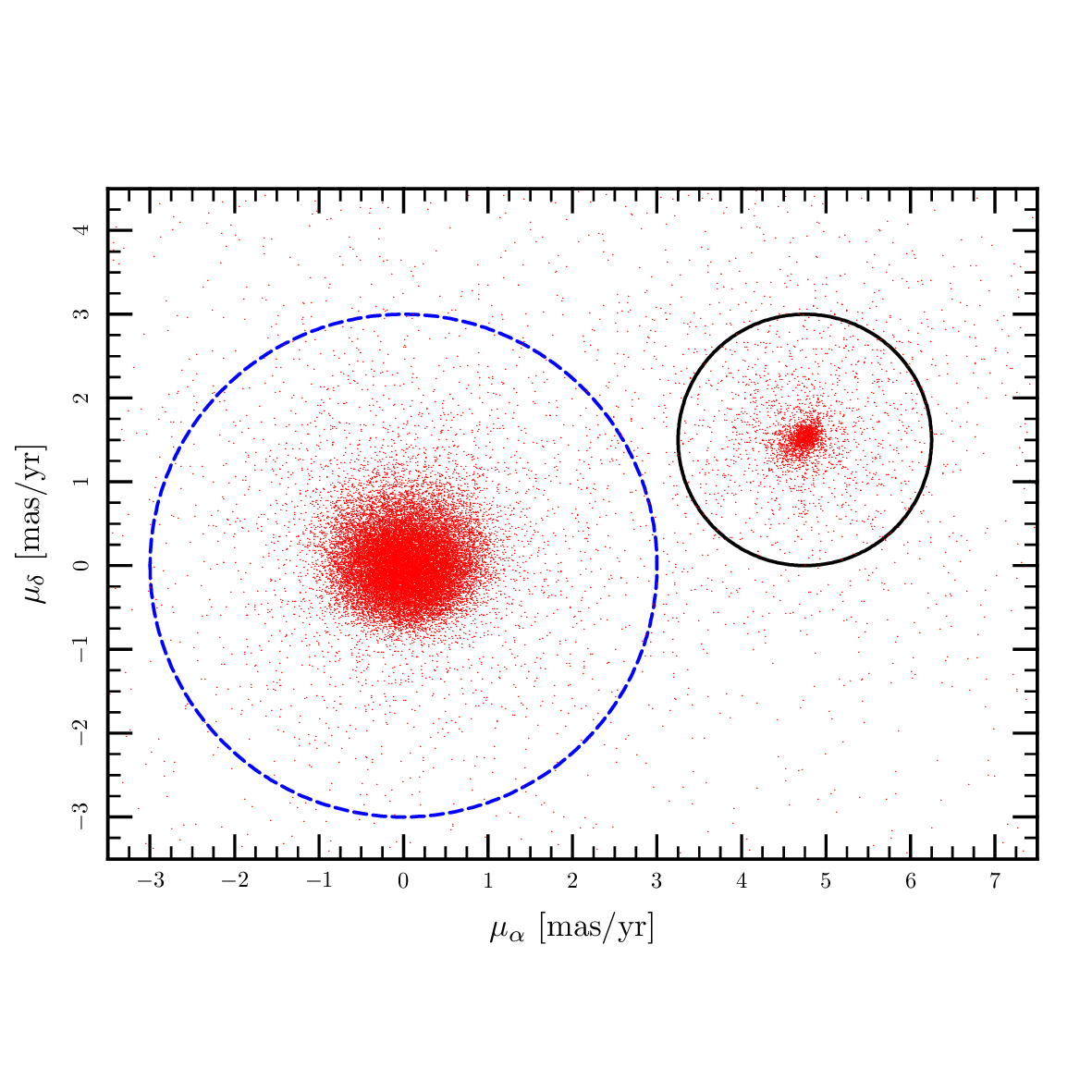,width=1.05\linewidth,clip=} 
\caption{PMs of all stars in our frames.   The large dashed blue
  circle encompasses the full 47 Tuc sample.  The solid black circle is the boundary of the PM sample for the SMC.  }
\label{fig:pmall}
\end{figure}

\begin{figure*}
\epsfig{file=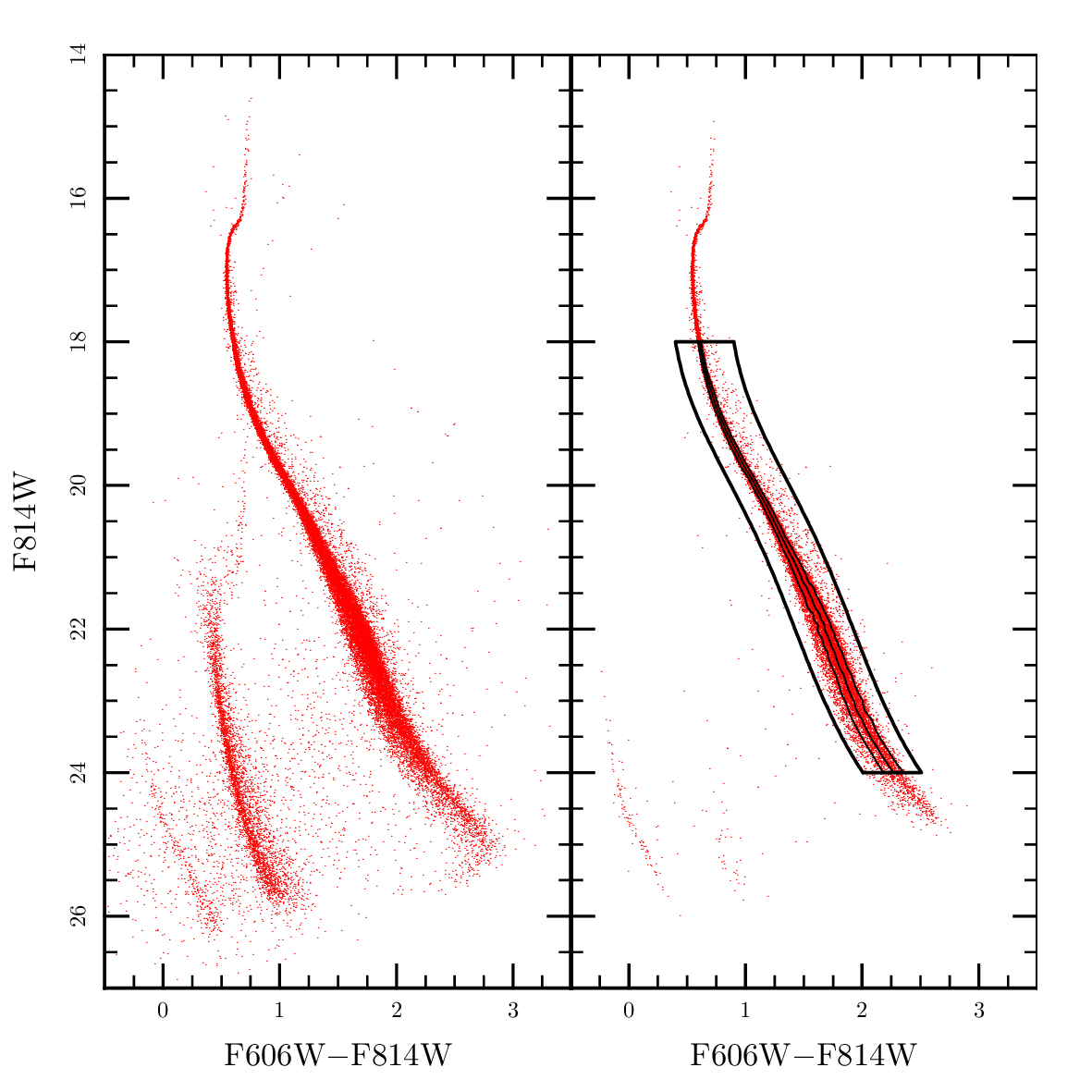,width=0.605\linewidth,clip=,bb=0 5 340 340} \hfill
\epsfig{file=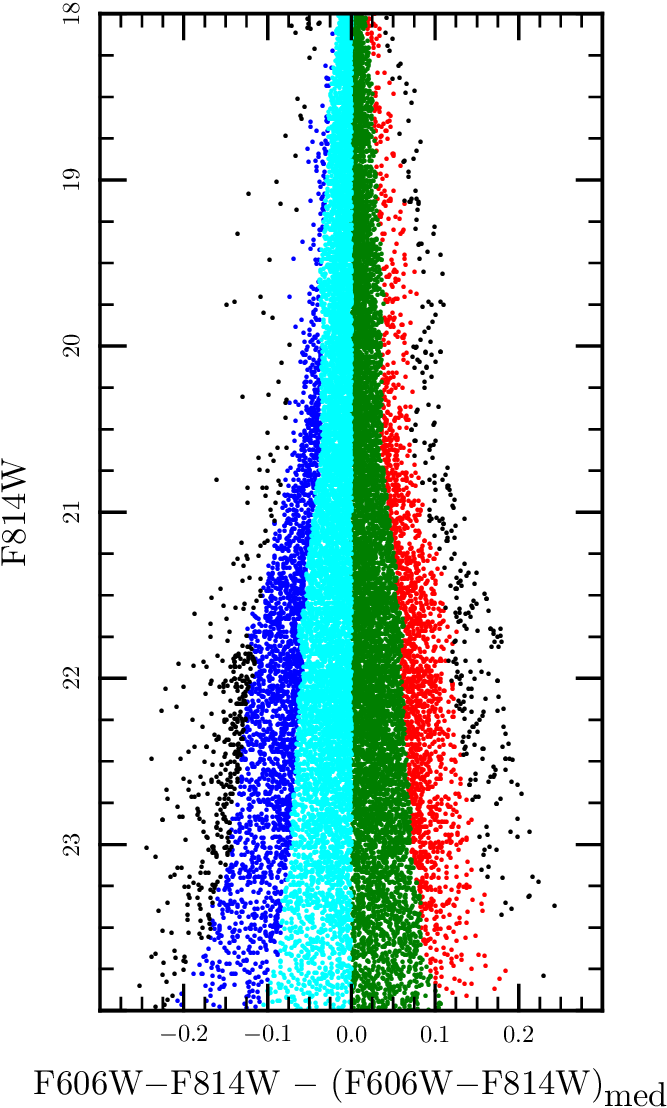,width=0.345\linewidth,clip=} 

\caption{CMDs of 47 Tuc. Left: All objects detected in our
 47 Tuc field (centered at 1.9 half light radii) with S/N $>$ 30.  Middle: The same data as in the left panel
 but now PM-cleaned to eliminate all but the 47 Tuc
 stars. Only the MS stars within the indicated box are
 retained for analysis. The light lines in this panel delineate
 the color groups outlined in \S~\ref{sec:prop-moti-kinem}. Right: The straightened CMD indicating the four color groups used in the analysis.
 The black points at the color extremes were not used - they were more than two standard deviations away from the median color at a given magnitude.
}
\label{fig:cmd}
\end{figure*}

Figure~\ref{fig:pmall} displays the PM diagram for all
stars in our frames.  The 47 Tuc PM distribution was centered around (0,0) with the distribution
centered near (4.5,1.5) for the SMC. We wanted to have as complete a sample as possible for both 47 Tuc and the SMC
with a minimum of contamination. For this reason our PM cut was very generous, about 10$\sigma$ in each case.
The left panel of Figure~\ref{fig:cmd} presents the cluster CMD using all the stars on the frames detected with S/N $>$ 30. We chose such a restrictive cut as our focus in this reduction is high-precision photometry and astrometry, so that is why the CMD 
 does not penetrate to very faint
magnitudes.   The central panel contains the PM-selected sample. This PM-cleaned CMD allows us to determine the
MS ridge line and, in addition, define a color and magnitude selected
sample of stars whose magnitudes lie within the black box.
 The solid black circle in
Figure~\ref{fig:pmall} is the boundary of the PM sample for the SMC.
We use both this PM selection and a color cut for the SMC sample in a similar manner.

%\begin{figure}[h]
% \epsfig{file=jay_cmd_paper.eps,width=1.05\linewidth,clip=} 
 %\caption{CMDs of 47 Tuc. Left: All objects measured in our
 % 47 Tuc field centered at 1.9 half light radii. Right: The same data as in the left panel
 % but now PM cleaned to eliminate all but the 47 Tuc
 % stars. Only the MS stars within the indicated box are
 % retained for analysis. The light lines in the right panel delineate
 % the color groups outlined in \S~\ref{sec:prop-moti-kinem}.
%}
%\label{fig:cmd}
%\end{figure}

\section{PM Kinematics}
\label{sec:prop-moti-kinem}

We divide the 47 Tuc MS into four groups in color from bluest to
reddest. We first sort the MS stars from brighest to faintest in F814W
and divide this sorted sample into thirty non-overlapping groups of
800 stars each.  For each group of stars we determine the median
F606W $-$ F814W color as well as the standard deviation of the
distribution as estimated by the $Q_n$ statistic (Rousseoeuw \& Croux
1991).  As in Heyl et al. (2012) we use this statistic as a robust
estimator of the standard deviation of a distribution, and we will
denote its value by $\hat \sigma$.  The right panel of
Figure~\ref{fig:cmd} displays the four color groups along the straightened MS of 47
Tuc as a function of apparent magnitude.

Thus we have defined the width through $\hat \sigma$ and the center of
the MS through the median as a function of F814W.  By
spline interpolation we determine the values of the width and median
at the value of F814W for each star in the sample and assign each star
to one of four color groups.  The first group lies greater than one
standard deviation and less than two standard deviations blueward of the median.  The second group lies
blueward of the median yet within one standard deviation.  The third
group lies redward of the median yet within one standard deviation.
The fourth group lies more than one standard deviation and less than two standard deviations redward of the
median.  The PMs of all of the stars lie within the large
circle in Figure~\ref{fig:pmall}, and their magnitudes lie within the
box in Figure~\ref{fig:cmd}.  The ridge line of the MS
typically lies within the second group.  The fourth group contains
some binary stars with nearly equal mass components.  Binaries with
more unequal mass components lie closer to the MS ridge
line.  Chemical abundance variations can also affect the color of
single stars resulting in a spread in color.  Our data are not
sufficient to determine whether a particular stellar object is a
single star with a peculiar chemical abundance, a binary or an optical
double.  Because the measured binary frequency for mass ratios greater than 0.6 in 47 Tuc is only a few percent (Milone et al. 2012b), 
the binaries will not significantly affect our results. These four color groups, indicated on the 47 Tuc CMD in the
middle panel of Figure~\ref{fig:cmd}, presumably mostly represent stars of
differing chemical composition with a small admixture of binaries.

Milone et al. (2012a) found in their analysis of the 47 Tuc MS that the various populations were often intertwined and crossed each other in the CMD.
This was generally seen in the most ultraviolet filters which are much more sensitive to chemical composition variations than F606W and F814W used here.
We tested whether He-enhanced MS stars at the metallicity of 47 Tuc
were always bluer in color than He-normal stars at all luminosities by overlaying a series of models (Dotter et al. 2008) with various enhanced He compositions  on the CMD. Over the entire extent
of the MS explored in this paper, the He-rich MS remained blue-ward of the He-normal sequence.  So we can be reasonably confident that the sequences of various abundances do not cross in the filters used here -- the
populations do not mix.

\begin{figure}
\epsfig{file=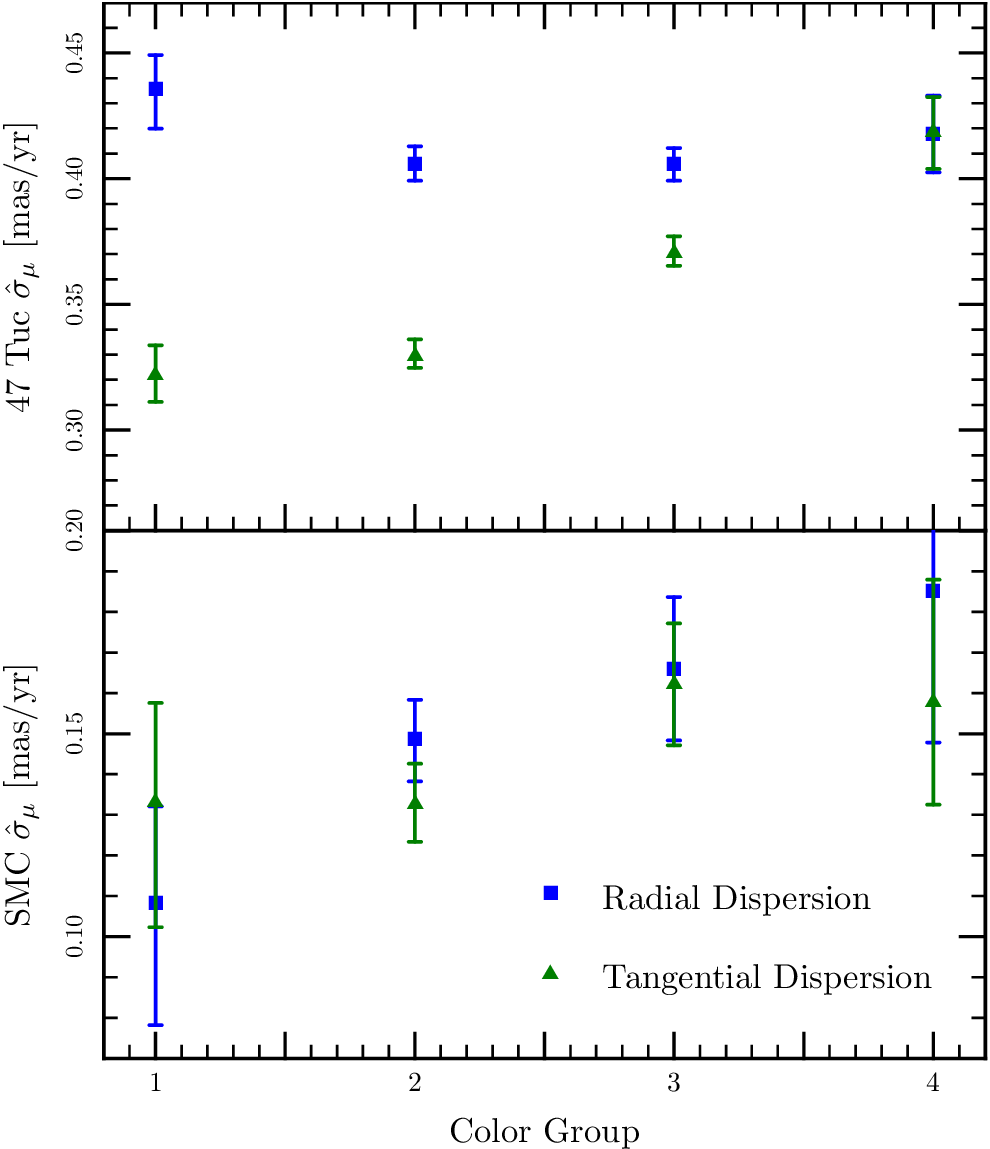,width=1.00\linewidth,clip=}
 \caption{The radial and tangential PM dispersions as a
   function of color group (1 is the bluest MS group, 4 the
      reddest) for 47 Tuc (top) and the SMC (bottom).} 
\label{fig:colorpms}
\end{figure}

In Figure~\ref{fig:colorpms} (top) we plot the PM dispersions in
the radial and tangential directions with respect to the cluster
center for the four groups.  In calculating these dispersions we have
added an additional cut in the estimated PM error.  We use only
 the stars whose errors are less than twice the median PM
error at their apparent magnitude.  The error bars in the dispersions
denote 90\% confidence regions as determined by bootstrapping the
sample. 

 If the stars are moving isotropically, the dispersions in
these two orthogonal directions should be the same. Clearly they are
not. The anisotropy is largest for the bluest stars and reduces to no
discernible anisotropy for the reddest ones.   

% \begin{figure}
% \epsfig{file=fourmedcoly.eps,width=1.05\linewidth,clip=}
% %\includegraphics[width=1.05\linewidth]{fourmedcol.eps}
%  \caption{The median PMs in the tangential
%    direction as a function of color group (1 is the bluest MS group, 4 the reddest).} 
% \label{fig:colormedpms}
% \end{figure}

\begin{figure}
\epsfig{file=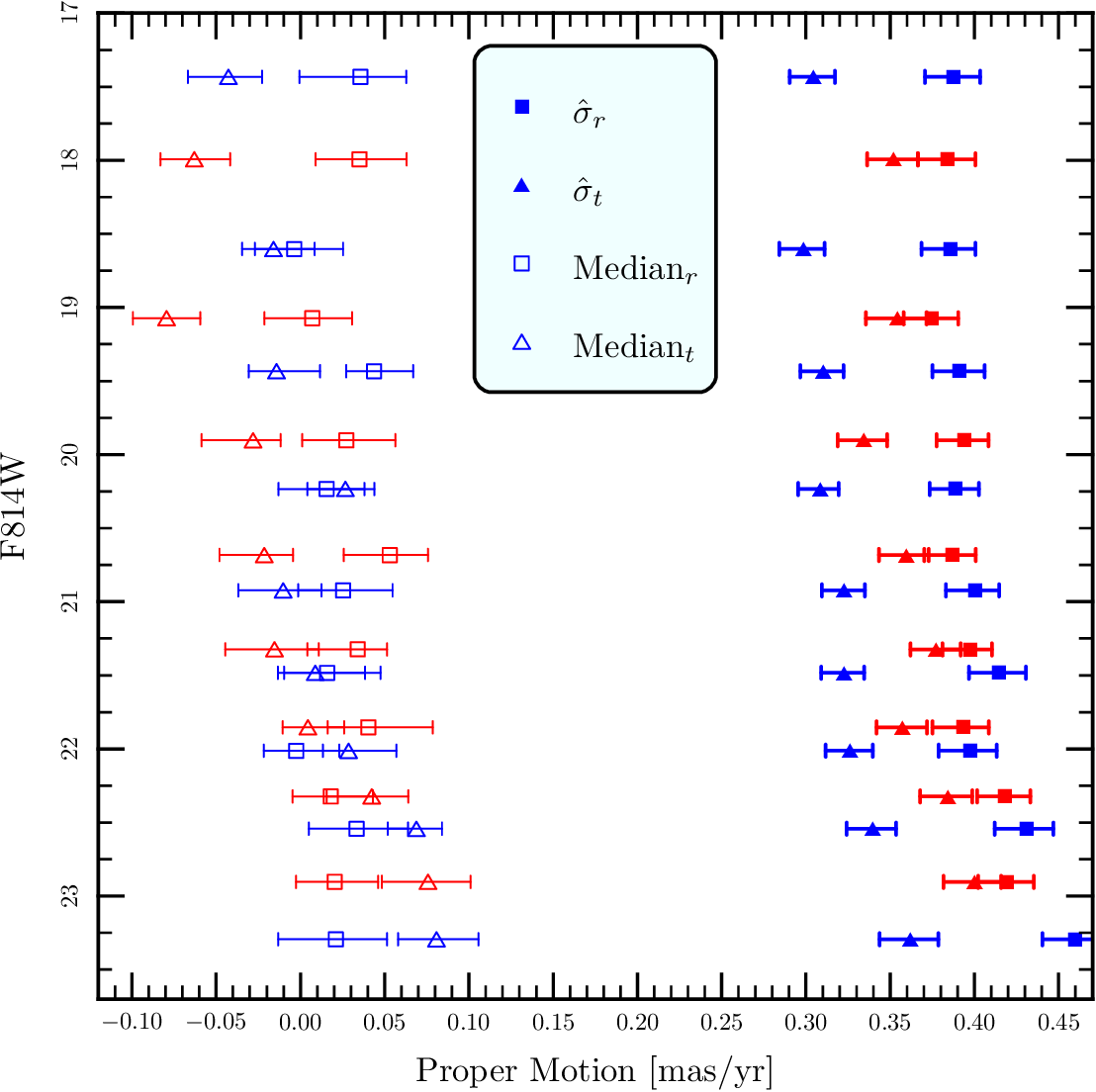,width=1.05\linewidth,clip=}
 \caption{The PM median values (open symbols) and  dispersions (filled symbols) in the tangential
   and radial directions as a function of magnitude for blue MS stars (groups 1 and 2 together) and the red MS stars (groups 3 and 4).} 
\label{fig:fourcol3b}
\end{figure}

This signature persists at all magnitudes. 
In Figure~4, we combine bins 1$+$2 and bins 3$+$4 and examine the dispersion of the blue and red stars as a function of magnitude, with 1000 stars in each magnitude bin.
At all magnitudes, the dispersion of the
blue stars in the tangential direction is always significantly smaller
than their radial dispersion while for the red stars the dispersions
exhibit little difference. The left-most set of points in this diagram are the median values of the proper motions of the blue and red stars for which there are no
discernible trends.
 
As a sanity check, we carried out a similar analysis on the SMC
stars. Again we divide the MS of the SMC into four color bins.  Of
course, since the MS of the SMC is much bluer than that of 47 Tuc, the
SMC groups have different color ranges compared to those in 47
Tuc.  We have restricted our sample here to lie below the turnoff of the
SMC (fainter than F814W$=22$) and brighter than 24 in F814W.
We measured the PM dispersions along the same radial and
tangential axes of the direction to the 47 Tuc center as a check on any potential systematic effects. Clearly
these axes have no physical significance for the SMC. These dispersions with color group are
illustrated in the bottom panel of Figure~\ref{fig:colorpms}.
 
The PMs and their dispersions here  are much smaller than in 47 Tuc reflecting
mainly the SMC's greater distance; it is twelve times farther from the Sun
than is 47 Tuc.  In contrast with 47 Tuc,
there is no evidence for any anisotropy in these PMs.  
  
\section{Radial Effects}

In addition to PM effects, we searched for radial differences among
the various 47 Tuc color groups. Since 47 Tuc is at best barely
relaxed, any discernible signals here could potentially provide
important clues to formation scenarios of the populations. 
From KS tests of significance on the cumulative radial distributions of the various MS color groups, we find that
the bluest group is less and less likely to be drawn from the same radial  distribution as the other groups 
as we progress red-ward. A comparison of group 1's with group 4's distribution yields a probability of only  2.80$\times10^{-4}$ that these two distributions were
drawn from the same parent sample. In these comparisons, the bluest group is always the most centrally concentrated. 
If the spread in MS color of 47 Tuc was due to a large
contribution of binaries, one would expect that the
reddest color bin would be the most centrally concentrated due to
mass segregation of these heavier stars.
 
  \section{Discussion}

The preceding analysis has shown 
that the 47 Tuc MS stars demonstrate anisotropic PMs that are strongly correlated with their colors. The sense
of this result is that the bluest stars possess the most anisotropic motions (larger radial than tangential dispersions) while the reddest stars exhibit no measurable anisotropy. 
The motions of the SMC stars  are completely consistent with no anisotropy along orthogonal axes towards and at right angles to the center of 47 Tuc.   
In addition, the bluer 47 Tuc MS stars are more centrally concentrated than the redder stars.

These differences contradict the notion that the globular cluster
formed monolithically.  Milone et al. (2012a) demonstrate that the blue
cohort exhibits CNO processing and therefore likely He
enrichment. This seems to firmly establish these stars as second generation. 

What do we expect for the motions and radial distribution of this second generation of stars? Near the half-mass radius,
47 Tuc has gone through only three half-mass relaxation times and our field is near this radius. If the cluster is not yet relaxed at this radius,
we would expect this second generation cohort to be more centrally concentrated, which it is. This is because these stars formed from dissipational
gas expelled from massive first generation stars. Their current orbits are more radial as they are in the process of relaxing via two-body interactions
to a distribution that is characteristic of their low masses - they are slowing diffusing outward (radially) to accomplish this.

The first generation stars exhibit no measurable anisotropy. This is not entirely easy to understand. Whether they underwent violent
relaxation before the second generation formed, or whether they still retain a memory of their initial collapse, they should still
be on moderately radial orbits. This can be seen in Figure 4-21 of Binney and Tremaine (1987) where stars outside the central core region
of a model cluster that underwent violent relaxation have anisotropic velocities at late times. However, the relaxation of the first generation of stars in 47 Tuc may be
 more complete because the initial state of the cluster may have been quite clumpy achieving more thorough relaxation (see e.g. Heyl et al. 1996) and not quasispherical as most simulations assume.

These broad-brushstroked descriptions are
illustrative only and not unique, but they do demonstrate how 
dynamical measurements could constrain the birth history of the
cluster.  Detailed numerical simulations including gas dynamics would
be required to get a clearer picture.

 \medskip

Based on observations with the NASA/ESA Hubble Space Telescope,
obtained at the Space Telescope Science Institute, which is operated
by the Association of Universities for Research in Astronomy, Inc.,
under NASA contract NAS5-26555. These observations are associated with
proposal GO-11677. Support for program GO-11677 was provided by NASA
through a grant from the Space Telescope Science Institute which is
operated by the Association of Universities for Research, Inc., under
NASA contract NAS5-26555. H.B.R. and J.H. are supported by grants from
The Natural Sciences and Engineering Research Council of Canada and by
the University of British Columbia.  J.A., J.S.K., R.M.R and
M.M.S. were funded by NASA. The authors benefited from useful
conversations with Brett Gladman.

%*************************REFERENCES*************************************
%**********************************************************************

\end{document}